\documentclass[preprint,showkeys,preprintnumbers,amsmath,amssymb,prb]{revtex4}
\usepackage{graphicx}
\usepackage{dcolumn}
\usepackage{bm}
\usepackage{pifont}

\begin{document}

\title{Deep level transient spectroscopy study for the development of ion-implanted silicon field-effect transistors for spin-dependent transport}

\author{B. C. Johnson\footnote{Email: johnsonb@unimelb.edu.au}}
\affiliation{Australian Research Council Centre of Excellence for Quantum Computer Technology, The University of Melbourne, Victoria 3010, Australia}
\author{L. H. Willems van Beveren}
\affiliation{Australian Research Council Centre of Excellence for Quantum Computer Technology, 
The University of New South Wales, Sydney 2052, Australia}
\author{E. Gauja}%
\affiliation{Australian Research Council Centre of Excellence for Quantum Computer Technology, 
The University of New South Wales, Sydney 2052, Australia}
\author{J. C. McCallum}%
\affiliation{Australian Research Council Centre of Excellence for Quantum Computer Technology, The University of Melbourne, Victoria 3010, Australia}

\begin{abstract}

A deep level transient spectroscopy (DLTS) study of defects created by low-fluence, low-energy ion implantation for development of ion-implanted silicon field-effect transistors for spin-dependent transport experiments is presented. Standard annealing strategies are considered to activate the implanted dopants and repair the implantation damage in test metal-oxide-semiconductor (MOS) capacitors. Fixed oxide charge, interface trapped charge and the role of minority carriers in DLTS are investigated. A furnace anneal at 950$\rm ^{o}$C was found to activate the dopants but did not repair the implantation damage as efficiently as a 1000$\rm ^{o}$C rapid thermal anneal.  No evidence of bulk traps was observed after either of these anneals. The ion-implanted spin-dependent transport device is shown to have expected characteristics using the processing strategy determined in this study.

\end{abstract}
\maketitle

\newpage

\section{Introduction}

Solid-state quantum computers require spin readout and control in order to transfer information to and from spin-encoded electrons or nuclei. A number of readout pathways have been identified based on spin-to-charge conversion.\cite{PhysRevB.69.233301,Elzerman:2004ig} Recently, spin dependent transport properties in a bulk-doped field-effect transistor (FET) observed by electrically detected magnetic resonance (EDMR) has been demonstrated and is therefore also a candidate for spin readout and control.\cite{2008ApPhL..93g2102W} In this work, a broadband on-chip terminated stripline is used to perform electron spin resonance on the hyperfine states of phosphorous donors in a silicon MOSFET (bulk doped). A second generation version of this device utilizing ion implantation technology may refine the MOSFET properties through the precision control of the depth, concentration and type of implanted dopants.

Ion-implantation is widely used in the fabrication of semiconductor devices. Spatial control is also easily achieved through the use of masks. High-quality oxide growth after implantation is problematic as implanted dopants will diffuse unless a dopant with a low diffusion coefficient is chosen such as Sb.\cite{lo:242106,schenkel:112101}  Alternatively, a low temperature high-quality oxide growth may proceed implantation to avoid any significant dopant diffusion.\cite{zhang:2964} In another approach, implantation can be performed through the oxide itself. Recently, implantation through a 5~nm oxide layer followed by a rapid thermal anneal (RTA) was shown not to increase the SiO$_{2}$/Si interface trap density and was in fact found to be beneficial when the as-grown oxide interface trap density is abnormally high.\cite{McCallum:2008nx} 

Deep level transient spectroscopy (DLTS) is well-suited to quantify the low defect concentrations   that can still be detrimental to device operation. In this work, measurements are presented for ion implanted MOS capacitors used to optimize the processing strategy of spin-dependent transport FETs. As implantation is used as any spurious signals arising from microwave absorption of the P donors in the bulk and in the P-diffused source and drain contacts in these FETs are avoided. The activation of donors as well as the minimization of oxide and interface traps are investigated. 

\section{Experiment}
Substrates for the test MOS capacitors were n-type, P-doped Si(100) Czochralski-grown wafers with a resistivity of 5-10~\underline{$\rm\Omega \cdot cm$}. The FETs described above were fabricated on an n-type, P-doped Si$^{28}$ (100) epilayer with a resistivity of 110~\underline{$\rm\Omega \cdot cm$}. Initially, a 200-300~nm thick uniform field oxide was thermally grown on both wafers in a wet O$_{2}$ ambient. Phosphorous in-diffusion in photolithographically etched back regions was then used to form the source and drain of the FET. A central window in the channel region of the FET and circular regions in the test wafer were etched back for growth of a 20~nm high-quality oxide grown using a triple wall furnace at 800~$\rm ^{o}C$ in a dry O$_{2}$ ambient. Dichloroethylene (DCE) was introduced during the growth to prevent ionic impurities being incorporated.

\begin{figure}[h]
 \begin{center}
 \rotatebox{0}{\includegraphics[height=14cm]{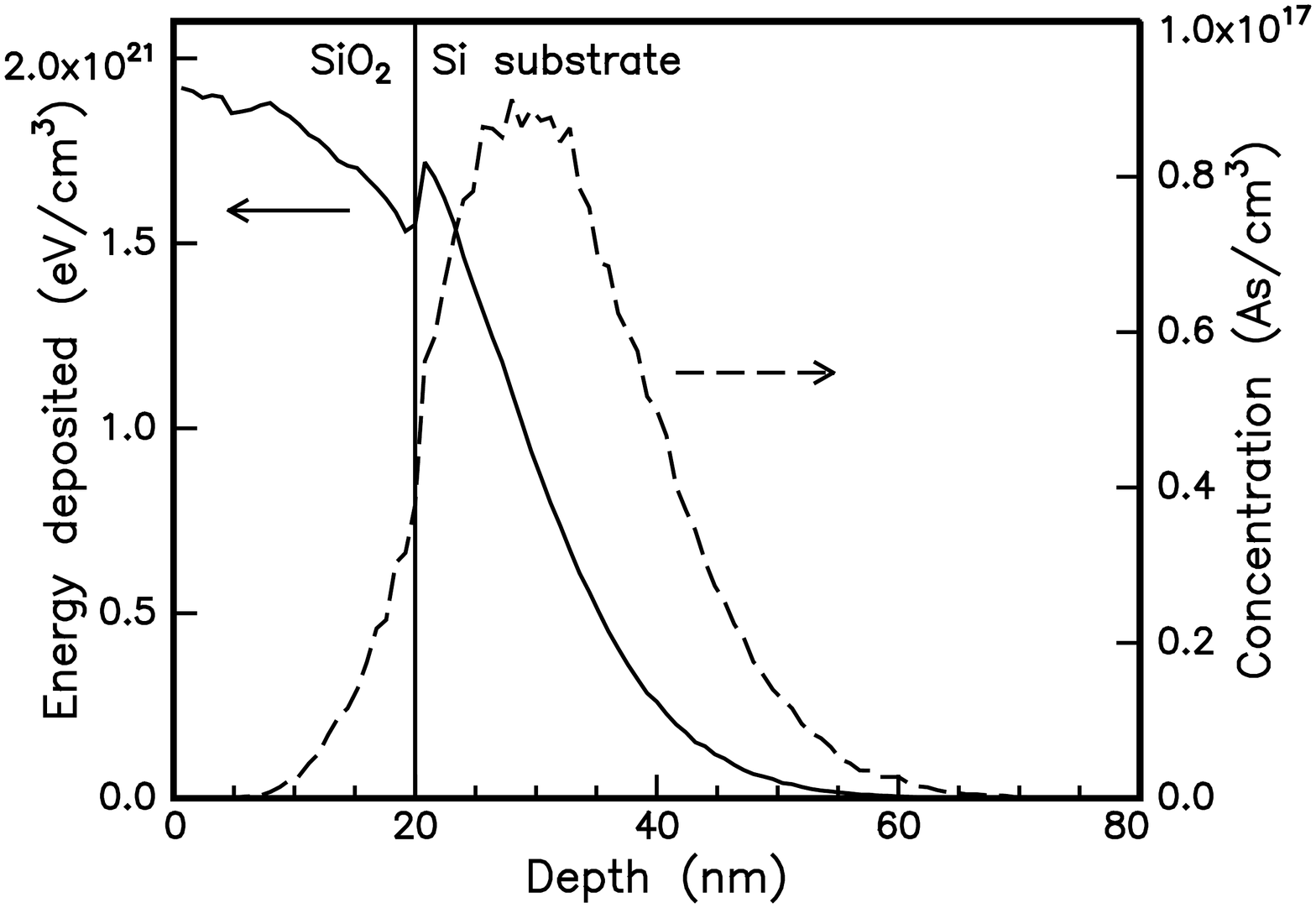}}
 \end{center}
\caption[]{Energy deposited and concentration profile simulation of 30 keV As to a fluence of $\rm2\times 10^{11}\; cm^{-2}$ using SRIM.\cite{Ziegler:1996lg} The Si/SiO$_{2}$ interface is indicated by the vertical line.}
\label{srim}
\end{figure}

Figure~\ref{srim} shows a simulation using SRIM\cite{Ziegler:1996lg} of the 30 keV As implant profile performed into the channel region of the FET and the circular regions of the test capacitors to a fluence of $\rm2\times 10^{11}\; cm^{-2}$. This implant gives a peak concentration of  $\rm0.9\times 10^{17}\; As/cm^{3}$ at a depth of 10~nm beyond the Si/SiO$_{2}$ interface, corresponding to a As donor spacing of about 20~nm. This is similar to the P spacing in the proposed solid-state quantum computer. \cite{Dzurak:2003fr} Most of the energy of the implant is deposited into the oxide. To repair the damage and activate the dopants various standard annealing strategies were considered. A rapid thermal anneal (RTA) at 1000~$\rm^{o}C$ for 5 seconds in a nitrogen ambient, a 950~$\rm^{o}C$, 30 minute furnace anneal (FA) in a forming gas ambient (95\% N$_{2}$, 5\% H$_{2}$) and a combination of RTA and FA (RTA+FA) were used. The diffusion length ($\sqrt{4D_{As}t}$) of the implanted As determined by the As diffusion coefficient, $D_{As}$, was found to be 1.8~nm and 16.3~nm for these RTA and FA anneals, respectively.\cite{Tsukamoto:1980tg} 

After the anneal, each sample underwent a passivation anneal in forming gas for 15 minutes at 400~$\rm^{o}C$. This anneal reduces the interface trap density by 20\%. Al was then evaporated to form the gate and ohmic contacts over the source and drain. The circular regions of the test wafer and the entire back surface was metallized with Al. Following metallization a further forming gas anneal was performed. 
 
Capacitance voltage (CV) and deep level transient spectroscopy (DLTS) were performed with a SULA Technologies DLTS system to characterize the MOS capacitors in a temperature range 80-370~K. The interface trap density was determined using the methodology outlined by Johnson and a capture cross section of \underline{$\rm 5\times 10^{-15} \; cm^{2}$} was assumed.\cite{Johnson:1982aa} 

\section{Results and discussion}

\begin{figure}[h]
 \begin{center}
 \rotatebox{0}{\includegraphics[height=14cm]{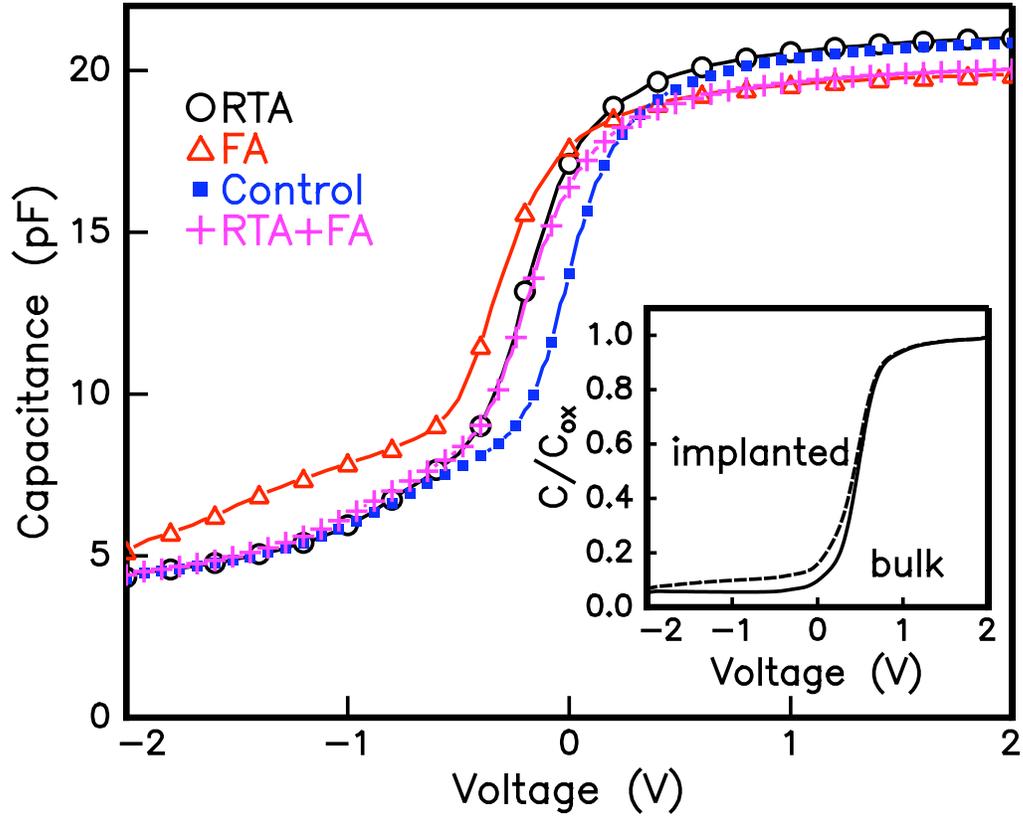}}
 \end{center}
\caption[]{Capacitance-voltage curves of processed MOS test capacitors measured with an AC frequency of 1~MHz. The measurement temperature was 300~K. The inset shows the simulated CV of a bulk doped (solid line) and As implanted wafer (dashed line).}
\label{cv}
\end{figure}

Figure~\ref{cv} shows the CV obtained from the MOS capacitors having undergone the three different annealing strategies. The CV from a control sample that had no implantation or high temperature anneal is shown for comparison. Samples that had undergone an RTA at some stage had similar CV profiles. The FA sample was shifted further to negative voltages and had a slightly higher inversion capacitance. All CV curves show a weak inversion response which is due to interface traps.\cite{Sedgwick:1968ys,Deal:1967ef} The presence of an activated As profile near the Si/SiO$_{2}$ interface will also increase the capacitance in weak inversion. In addition, the active donors will shift the flat-band voltage since the metal-semiconductor work-function difference will vary with shifts in the Fermi level in the semiconductor. Likewise, fixed oxide charge deposited into the oxide by implantation also shifts the flat-band voltage down. To qualitatively de-convolute these contributions we simulated the CV profiles using TCAD\cite{tcad} as shown in the inset of Fig.~\ref{cv}. No defects in the oxide or at the Si/SiO$_{2}$ interface were included in these simulations so that only changes due to the As implant could be observed. The simulation shows that the implantation does not shift the curve to lower voltages by any appreciable amount. Therefore, a large component of the shift in these experimental CV curves is attributed to fixed oxide charge. Although the FA is most efficient at activating the dopants indicated by the increase in the inversion capacitance, a comparatively large amount of fixed oxide charge remains after the anneal. The fixed oxide charge is estimated to increase by $\rm 2.9\times 10^{11}/cm^{2}$ whereas the RTA sample results in an increase of $\rm 1.6\times 10^{11}/cm^{2}$. This is calculated assuming that the shift is entirely due to the fixed oxide charge and therefore represents an upper limit.

\begin{figure}[h]
\begin{center}
 \rotatebox{0}{\includegraphics[height=14cm]{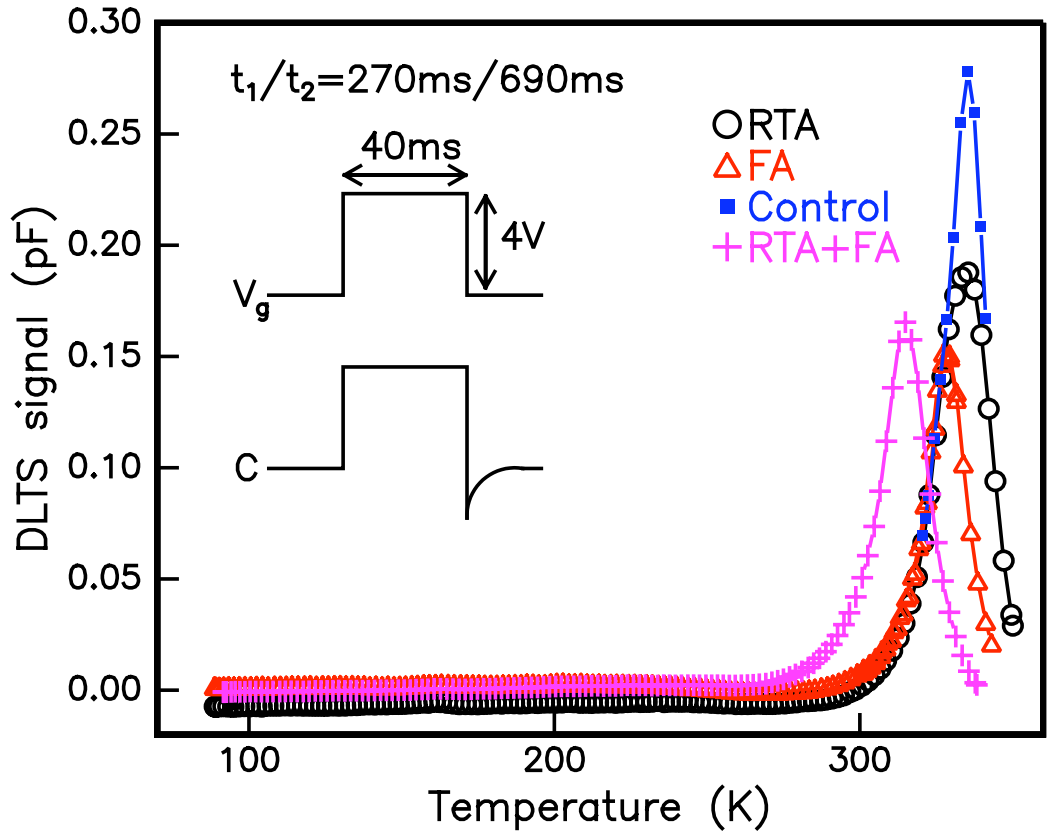}}
 \end{center}
\caption[]{DLTS signal as a function of temperature with the correlator time window defined by $t_{1}=270$~ms and $t_{2}=690$~ms. The voltage pulse (from -2V to 2V) and corresponding capacitance transient is also shown.}
\label{dlts}
\end{figure}

Figure~\ref{dlts} shows the DLTS correlator signals as a function of temperature. A continuous background as well as a large peak near room temperature is observed in all samples. These peaks are associated with minority carrier generation-recombination processes.\cite{Schulz:1977aa,Beyer:1999bh} These will be discussed further below. No peaks associated with bulk traps produced by the implantation were observed after any of the anneals. The interface trap density, $D_{it}$, calculated from the low temperature part of the spectra where emission from majority carriers dominates was $\rm1.4\times 10^{10}\; /eV/cm^{2}$ for the control sample. The RTA and (RTA+FA) both retained a low $D_{it}$ of $\rm1.3\times 10^{10}\; /eV/cm^{2}$. The FA had a slightly higher $D_{it}$ of $\rm 1.9\times 10^{10}\; /eV/cm^{2}$. These values are the densities of interface traps in the band gap at an energy of $(E_{c}-E_{t})=0.16$~eV.

The large peaks in Fig.~\ref{dlts} are due \underline{to generation-recombination processes through} \underline{interface minority carrier traps near the centre of the band gap.} The activation energy and capture cross section of \underline{this process} can be determined using an Arrhenius plot of $e_{p}/T^{2}$ versus $1/kT$ according to the hole emission rate equation\cite{book:sze}

\begin{equation}
\label{emission}
e_{p}=\sigma_{p}v_{th}N_{V}\exp\bigl( \frac{E_{f}-E_{V}}{kT}\bigl)=\sigma_{p}\gamma_{p}T^{2}\exp\bigl( \frac{\Delta E_{h}}{kT}\bigl)
\end{equation}

\noindent where $v_{th}$ and $N_{V}$ are the thermal velocity and the effective density of states, respectively. $\Delta E_{h}$ is the activation energy, $\sigma_{p}$ the \underline{capture cross section} and $\gamma_{p}$ is a constant. In Fig.~\ref{dlts}, the peaks near room temperature have a hole emission rate of 2.18~s$^{-1}$ determined by the sampling time $t_{1}$ and  $t_{2}$ and the equation $e_{p}={ln(t_{2}/t_{1})}/(t_{2}-t_{1})$. Using this equation and Eq.~\ref{emission} with various sampling times, $\sigma_{p}$ and $\Delta E_{h}$ can be determined.

\begin{figure}[h!]
 \begin{center}
 \rotatebox{0}{\includegraphics[height=14cm]{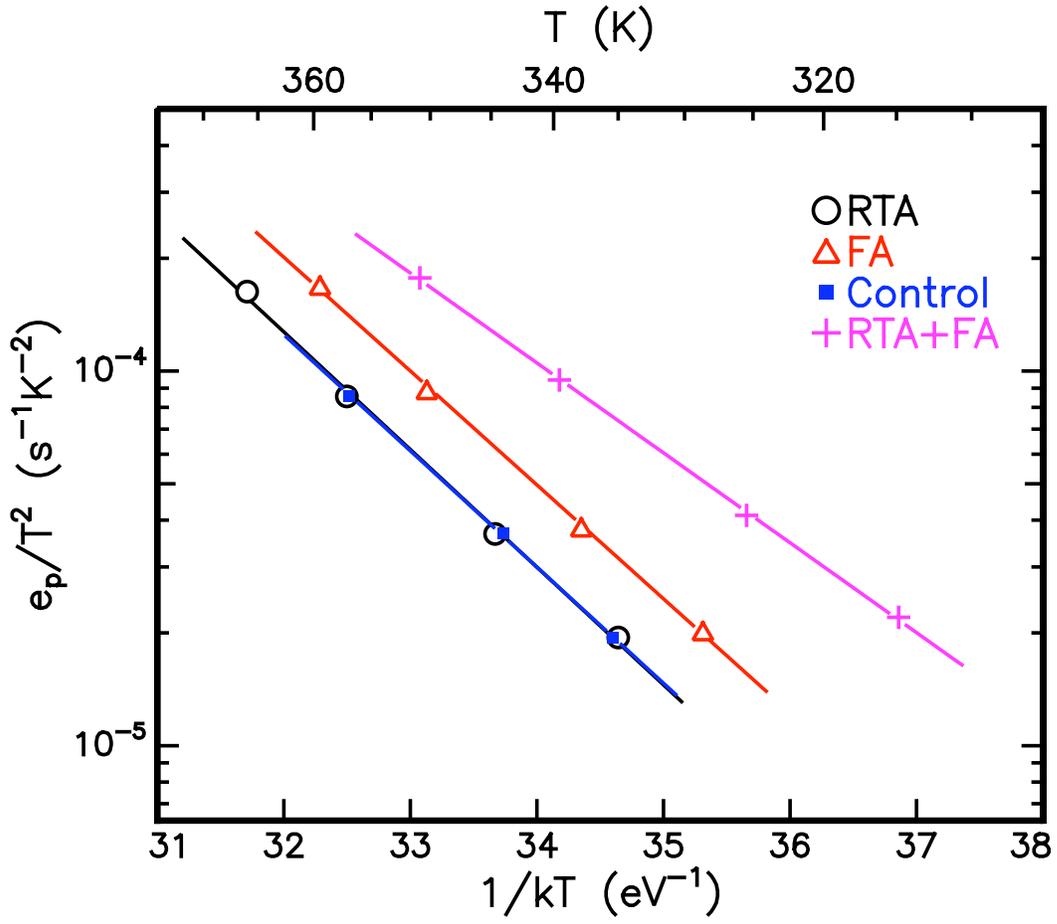}}
 \end{center}
\caption[]{Arrhenius plot of minority carrier peaks in the DLTS signal shown in Fig.~\ref{dlts} for each annealing strategy.
}
\label{arrhenius}
\end{figure}

Figure~\ref{arrhenius} shows the Arrhenius plots. The control, RTA and FA samples resulted in values within error of each other with $\Delta E_{h}=0.70$~eV and $\sigma_{p}=\underline{4.0\times 10^{-16}}$~cm$^{2}$. \underline{In contrast, the (RTA+FA) sample gave a $\Delta E_{h}$ and $\sigma_{p}$ of 0.55~eV and $5.18\times 10^{-18}$~cm$^{2}$,} respectively. \underline{Here, a different region of the continuous hole trap distribution is dominating} \underline{the generation-recombination process.} Since the room temperature peaks for the control, RTA and FA samples \underline{gave similar activation energies,}  their intensities which are proportional to the hole concentration at 0.7~eV above the valance band can be compared. The trap concentration decreased by 33\% after the RTA anneal and 46\% after the FA anneal.

\begin{figure}[h!]
 \begin{center}
 \rotatebox{0}{\includegraphics[height=14cm]{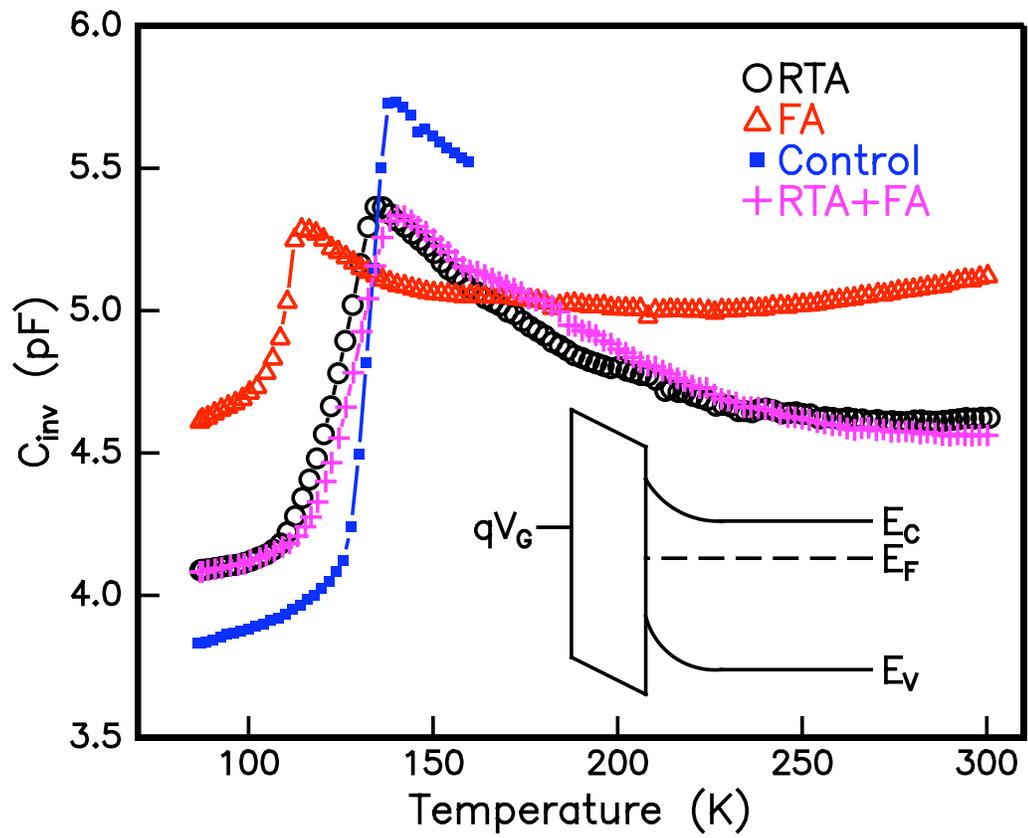}}
 \end{center}
\caption[]{Inversion capacitance at -2~V as a function of temperature (no pulse). The inset is a schematic of the band structure of the MOS capacitor during the temperature scan.
}
\label{cT}
\end{figure}

Evidence of minority carriers was also found in the inversion capacitance temperature scan as shown in Fig.~\ref{cT}. This scan is used to normalize the DLTS signals. A peak is observed at a temperature where an inversion layer is formed. The position of the peak is found to be dependent on the gate bias with -4~V shifting the peak 30~K higher. At temperatures above this peak the capacitance transient is dominated by minority carrier generation-recombination. All peaks occur at around 135~K which corresponds to an electron trap energy of $E_{C}-E_{T}=0.27$~eV. The FA sample has a slightly lower peak temperature giving $E_{C}-E_{T}=0.22$~eV. At energies further from $E_{C}$, $D_{it}$ cannot be determined since the capacitance transient is dominated by minority carriers. 

An investigation of enhanced minority carrier generation is underway. Preliminary results suggest that thinning of the low-quality field oxide during processing is the cause. A field produced by the gate shifts the Fermi level to a position where hole generation is more likely (to be published elsewhere). The role of minority carrier traps at the low temperatures ($<$4~K) where FET measurements are made is not yet known.

\begin{figure}[h!]
 \begin{center}
 \rotatebox{0}{\includegraphics[height=14cm]{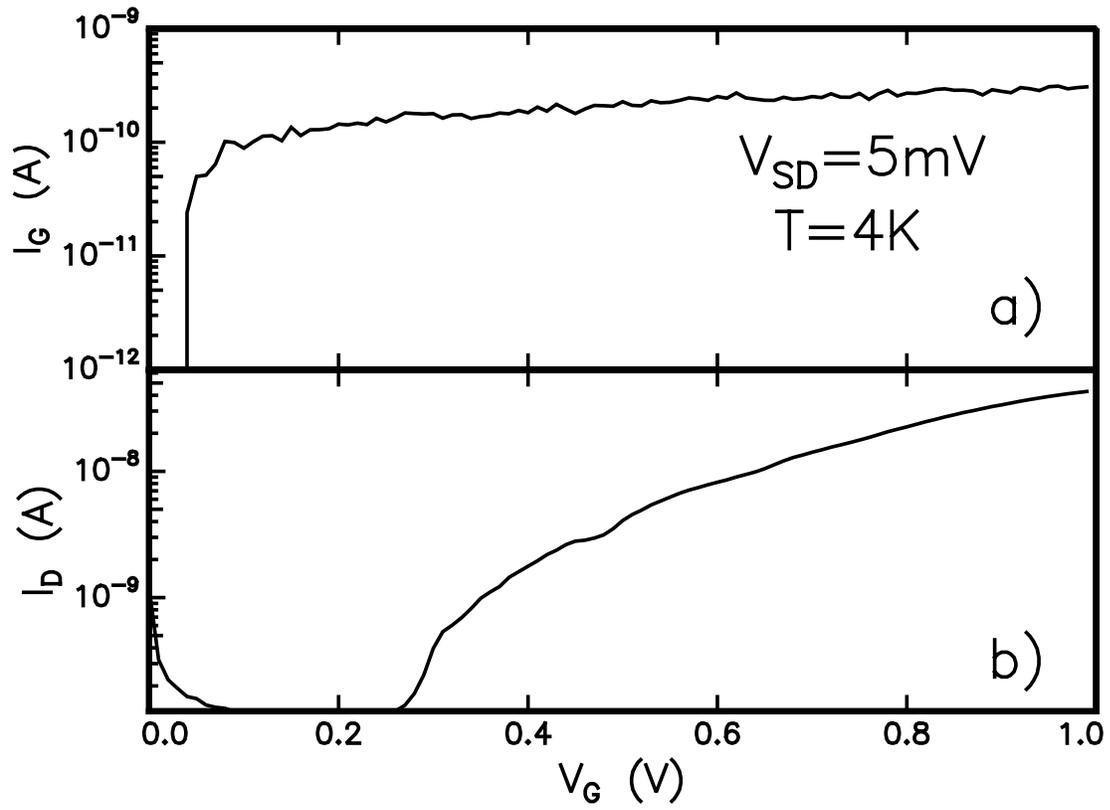}}
 \end{center}
\caption[]{a) Gate leakage and b) the drain current with V$_{SD}=+5$~mV as a function of gate voltage showing turn on of the FET device at 4~K.
}
\label{leakage}
\end{figure}

In order to ensure that there would be no gate leakage current in the FET device, an RTA was used. This was followed by the passivation anneal, metallization and post-metalization anneal. Figure~\ref{leakage} shows the gate leakage and the drain current as a function of gate bias for such a device. The trends between 0-0.1~V are features of the measurement apparatus and do not reflect the true current behaviour. The FET turn-on voltage compares well with the bulk doped version of the FET. These characteristics show that there are no serious issues preventing full device operation. EDMR measurements of this first As implanted spin-dependent transport FET are in progress.

\section{Conclusion}

The RTA after implantation resulted in the lowest fixed oxide charge and interface trap densities while also limiting the As diffusion. The capacitance transients in the DLTS spectra were found to be dominated by minority carrier generation-recombination processes at an unusually low temperature. It is suggested that a thin field oxide can act as a source of minority carriers. Although the DLTS peaks were only observed near room temperature, minority carrier processes dominated the inversion capacitance down to 130~K.

An RTA anneal was chosen for the fabrication of the FET device which showed the expected turn-on behaviour and minimal gate leakage. This suggests that implantation through an oxide layer followed by an RTA is a viable and realistic means of successfully fabricating spin dependent transport FETs.

Implanted Hall bar MOSFETs are now being developed to further study mobility and activation of implanted donors in the low-energy, low-fluence regime. The interaction of dopants with our high-quality  SiO$_{2}$/Si interfaces is of particular interest and importance.

\begin{acknowledgments}
The Department of Electronic Materials Engineering at the Australian National University is acknowledged for their support by providing access to ion implanting facilities. This work was supported by the Australian Research Council through the Centre of Excellence for Quantum Computer Technology and by the U.S. National Security Agency and U.S. Army Research Office (under Contract No. W911NF-04-1-0290). 
\end{acknowledgments}


\newpage

\end{document}